\documentclass[a4paper,UKenglish,cleveref, autoref, thm-restate]{lipics-v2021}

\usepackage{algorithm}
\usepackage{algpseudocode}
\usepackage[beginComment=//~,beginLComment=//~,endLComment=~,]{algpseudocodex}

\newcommand{\tool}{\textsc{DSA }}
\newcommand{\ds}{D-strings }

 \newcommand{\defproblem}[3]{
   \vspace{2mm}
 \noindent\fbox{
   \begin{minipage}{0.96\textwidth}
   \textsc{#1}\\
   {\bf{Input:}} #2  \\
   {\bf{Output:}} #3
   \end{minipage}
   }
   \vspace{2mm}
 }

\title{Fast Exact String to D-Texts Alignments}

\author{Njagi Moses Mwaniki}{Department of Computer Science, University of Pisa, Italy}{njagi.mwaniki@di.unipi.it}{}{}

\author{Erik Garrison}{University of Tennessee Health Science Center, USA}{egarris5@uthsc.edu}{}{}

\author{Nadia Pisanti}{Department of Computer Science, University of Pisa, Italy}{nadia.pisanti@unipi.it}{https://orcid.org/0000-0003-3915-7665}{}

\authorrunning{Mwaniki, Garrison, Pisanti} 

\Copyright{Njagi Moses Mwaniki and Erik Garrison and Nadia Pisanti} 



\funding{\newline
\begin{minipage}{0.1\textwidth}
\includegraphics[height=1cm]{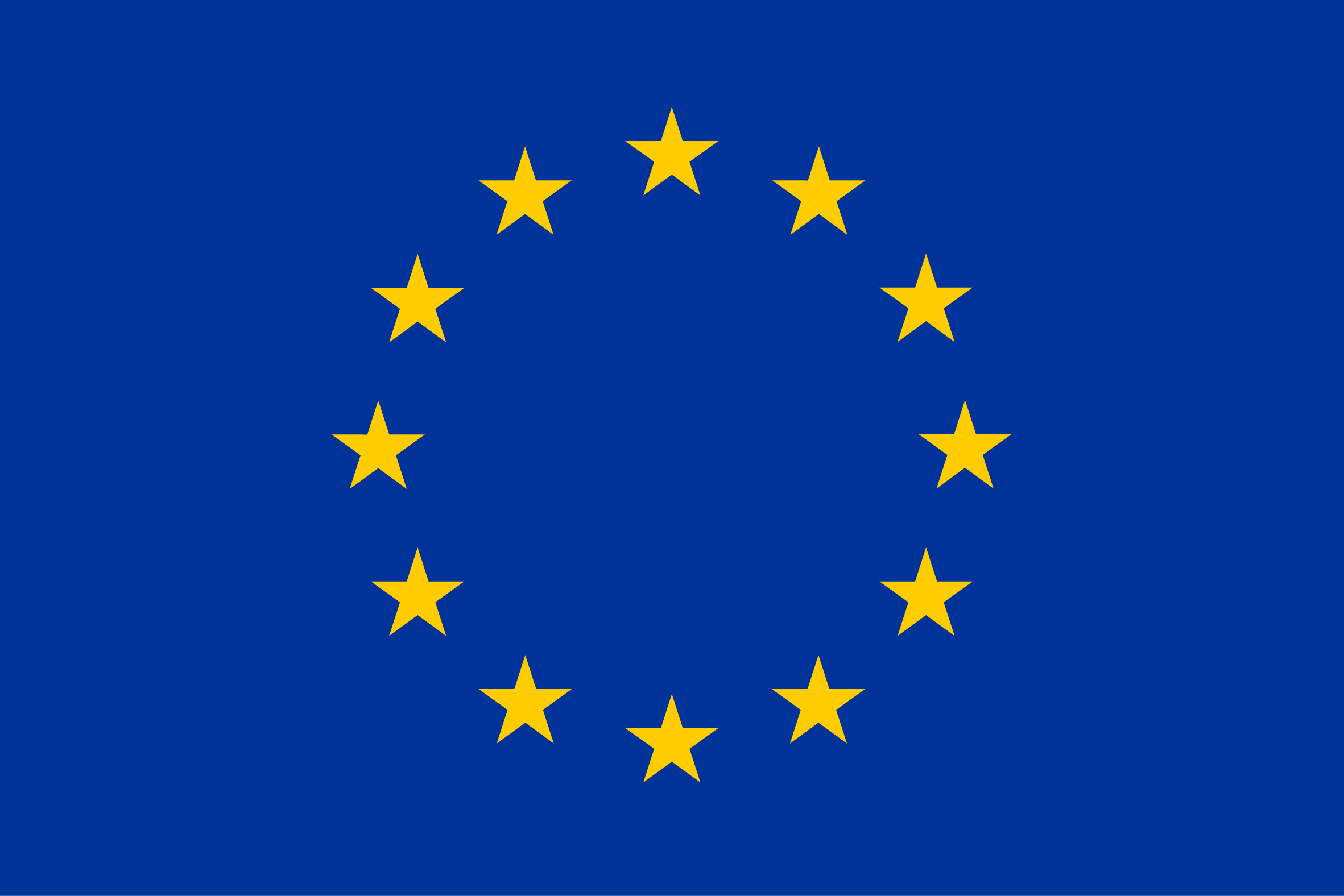}
\end{minipage}
\hfill
\begin{minipage}{0.88\textwidth}
This paper is part of a project that has received funding from the European Union’s Horizon 2020 research and innovation programme under the Marie Skłodowska-Curie grant agreement No 956229. Co-financed by the Connecting Europe Facility of the European Union.
\end{minipage}
\smallskip
}

\category{} 

\relatedversion{} 

\nolinenumbers 

\EventEditors{John Q. Open and Joan R. Access}
\EventNoEds{2}
\EventLongTitle{42nd Conference on Very Important Topics (CVIT 2016)}
\EventShortTitle{CVIT 2016}
\EventAcronym{CVIT}
\EventYear{2016}
\EventDate{December 24--27, 2016}
\EventLocation{Little Whinging, United Kingdom}
\EventLogo{}
\SeriesVolume{42}
\ArticleNo{XX}

\begin{document}

\maketitle

\begin{abstract}
In recent years, aligning a sequence to a pangenome has become a central problem in genomics and pangenomics. A fast and accurate solution to this problem can serve as a toolkit to many crucial tasks such as read-correction, Multiple Sequences Alignment (MSA), genome assemblies, variant calling, just to name a few.
In this paper we propose a new, fast and exact method to align a string to a D-string, the latter possibly representing an MSA, a pan-genome or a partial assembly. \\
An implementation of our tool dsa is publicly available at \url{https://github.com/urbanslug/dsa}.
\end{abstract}

\section{Introduction}
\label{sec:intro}

In recent years, aligning a sequence to a pangenome has become a central problem in genomics and pangenomics (\cite{consortium_computational_2018}). This problem has often been addressed with the purpose of aligning sequencing reads to a complex structure such as a degenerate string or a more general genome graph structure (e.g.\cite{garrison2018,vg2021,vargas,DBLP:journals/bioinformatics/RautiainenMM19,graphaligner,minigraph}), and hence assuming the query string to be considerably shorter than the pangenome (e.g. \cite{garrison2018,vg2021,cislak_sopang_2018,cislak_sopang_2020,vargas}). As sequencing data is recently offering longer sequences with high accuracy, the attention has very recently been moved towards the problem of aligning to a graph-like structure a string which is approximately as long: a fast and accurate solution to this problem would serve as a toolkit to many crucial tasks such as read-correction, Multiple Sequences Alignment (MSA), genome assemblies, variant calling, just to name a few. To this purpose, because the requirement is an alignment that is at the same time exact and sensible to indels as well as mismatches, then new computational challenges are faced: current exact methods that perform exact alignments to long strings would have a quadratic complexity with high memory requirements, and hence be prohibitively slow. \\
Consider the following MSA of three closely-related sequences (on the left) and its compact representation as a D-string (on the right):\\
     \begin{minipage}{0.43\textwidth}
\begin{tabular}{ll}
& \texttt{GCA{\color{red}AT}C{\color{red}G}GG{\color{red}TA}TT}\\
& \texttt{GCA{\color{red}AT}C{\color{red}G}GG{\color{red}AA}TT}\\
& \texttt{GCA{\color{red}CG}C{\color{red}T}GG{\color{red}AT}TT}
\end{tabular}
    \end{minipage}
    \begin{minipage}{0.57\textwidth}
\[ 
\hat{T}=
  \texttt{GCA} 
 [ {\color{red}
  \texttt{AT} / 
  \texttt{CG} }
]
  \texttt{C} 
\left  [ {\color{red}
  \texttt{G}/
  \texttt{T} }
\right ]
  \texttt{GG} 
  [ {\color{red}
  \texttt{TA}/
  \texttt{AA}/
  \texttt{AT} }
] 
\texttt{TT}
\]
   \end{minipage}

The D-string (or D-text) $\hat{T}$ contains some deterministic (shown in black) and some non-deterministic (shown in red) segments. Formally, $\hat{T}$ is a sequence of n sets of strings
where the i$^{th}$ set contains strings of the same length $\ell_i$ (possibly $=1$ in the deterministic case) but this length can vary between different sets.\\
An {\em indeterminate string} over an alphabet $\Sigma$ is a sequence of subsets of $\Sigma$, and it basically corresponds to the special case of D-string in which $\ell_i=1$ for all $1\leq i\leq n$. A great deal of research has been conducted in the bioinformatic literature on this type of uncertain sequence (see~\cite{KMRC,cpm2005,KMRelat,Abrahamson:1987:GSM:37185.37191,DBLP:conf/isaac/CrochemoreIKRRW14,iliopoulos_et_al:LIPIcs:2016:6084} and references therein), that are equivalent to a sequence written in the IUPAC notation~\cite{IUPAC} to represent a position in a DNA sequence that can have multiple possible alternatives. These are commonly used to encode the consensus of a population of sequences~\cite{filter08,consortium_computational_2018,alzamel_degenerate_2018,alzamel_comparing_2020,IndetCPM20} in a multiple sequence alignment (MSA). \\
The more general notion of ED-strings (where within a degenerate position variants can have different sizes), and over them the short read matching problem {\em elastic-degenerate string matching} (EDSM) problem has attracted some attention in the combinatorial pattern matching community. Since its introduction in 2017~\cite{iliopoulosLATA2017}, a series of results have been published both for the exact (\cite{grossi_-line_2017,bernardini_even_2019,aoyama_faster_2018,sicomp}) as well as for the approximate (\cite{bernardini_approximate_2020,spire17}) version of the problem. \\
In this paper we propose a new, fast and exact method to align a string to a D-string, the latter possibly representing an MSA, a pan-genome or a partial assembly. Ours is a base-level heuristic-free alignment method that allows affine gap penalty score functions and the use of arbitrary scores as well as weights. \ds are one out of many others possible pangenome representation (\cite{graphrep}). Our algorithm exploits the assumption that the similarity the alignment must detect is high enough to suitably adapt the seminal idea of \cite{myers_anond_1986} to D-strings, combining at the same time \emph{partial order alignments} (\cite{lee_multiple_2002}) and the \emph{wavefront} paradigm (\cite{marco-sola_fast_2020}) managing to work with affine gap penalty function to score gaps, which is a desirable property when the target of the alignment is to account for INDELs variant events.

This paper is organised as follows: Section~\ref{sec:prel} gives some preliminary definitions on D-strings and notions of partial order  and wavefront alignment, Section~\ref{sec:algo} describes our algorithm, and Section~\ref{sec:exp} the experimental validation of the resulting tool dsa we realized.\\
The implementation of \tool is publicly available at 
\url{https://github.com/urbanslug/dsa}, and our \ds random generator can be found at \url{https://github.com/urbanslug/simed/}.

\section{Preliminary Notions}
\label{sec:prel}

\subsection{D-strings}
\label{ssec:ds}

A \textit{string} $X$ is a sequence of elements on an alphabet $\Sigma$, where the \textit{alphabet} $\Sigma$ is a non-empty finite set of letters of size $|\Sigma|$. The set of all finite strings over an alphabet $\Sigma$, including the \textit{empty string} $\varepsilon$ of length $0$, is denoted by $\Sigma^*$, while $\Sigma^+$ denotes the set $\Sigma^* \setminus \{\epsilon\}$. The set of all strings of length $k\!>\!0$ over $\Sigma$ is denoted by $\Sigma^k$. For any string $X$, we denote by $X[i, j]$ the \textit{substring} of $X$ that {\em starts} at position $i$ and {\em ends} at position $j$. In particular, $X[0 , j]$ is the \textit{prefix} of $X$ that ends at position $j$, and $X[i , |X|]$ is the \textit{suffix} of $X$ that starts at position $i$, where $|X|$ denotes the \textit{length} of $X$. 

\begin{definition}
A \emph{degenerate string (D-string)} $\hat{S}=\hat{S}\{1\}\hat{S}\{2\}\dots\hat{S}\{n\}$ of {\em length} $n$ over an alphabet $\Sigma$ is a finite sequence of $n$ degenerate letters $\hat{S}\{i\}$'s. Each \textit{degenerate letter} $\hat{S}\{i\}$ has \emph{width} $\ell_i\!>\!0$ and is a finite non-empty set of $|\hat{S}\{i\}|$ strings of the same length $\ell_i$ (i.e. $\hat{S}\{i\}[1],\ldots, \hat{S}\{i\}[|\hat{S}\{i\}|]  \in \Sigma^{\ell_i}$). 
\end{definition}

For any $1\leq i \leq n$, we remark that a degenerate letter $\hat{S}\{i\}$ with $|\hat{S}\{i\}|=\ell_i=1$ is just a simple letter of $\Sigma$. Whenever this happens, 
we will say that $i$ is a \emph{solid position} of $\hat{S}$. We now define some parameters that measure the degeneracy of a D-string.

\begin{definition}
The \emph{total size} $N$ and \emph{total width} $w(\hat{T})=W$ of a D-string $\hat{S}$ are respectively defined as $N=\sum_{i=1}^{n}|\hat{S}\{i\}| \cdot \ell_i$ and $W=\sum_{i=1}^{n}\ell_i$.
\end{definition}

\begin{example}
In the D-string $\hat{T}=\texttt{GCA[AT/CG]C[G/T]GG[TA/AA/AT]TT}$, $\ell_4=\ell_9=2$ while all other $\ell_i$'s are $1$, and  $|\hat{S}\{4\}|=|\hat{S}\{6\}|=2$, $|\hat{S}\{9\}|=3$ while all the other $|\hat{S}\{i\}|$'s are $1$; finally, the solid positions of $\hat{T}$ are $1,2,3,5,7,8,10,11$.
\end{example}

We remark that a D-string of width $W$ actually represents a set of linear strings of length $W$, corresponding to  all the strings that can be read by making any choice at a degenerate positions. Formally, given a D-string $\hat{T}$ of width $W$, and a string $T$ with $|T|=W$ being any string $T$ in such set, we say that $T$ belongs to $\hat{T}$ ($T\in \hat{T}$).

\begin{example}
The strings $\texttt{GCACGCTGGAATT}$ and $\texttt{GCAATCTGGTATT}$ are two of the twelve strings that belong to the D-string $\hat{T}=\texttt{GCA[AT/CG]C[G/T]GG[TA/AA/AT]TT}$.
\end{example}

The $i^{th}$ position $\hat{S}\{i\}$ of a D-string $\hat{S}$ ($1\leq i \leq n$) should not be confused with its $i^{th}$ width $\hat{S}[i]$ ($1\leq i \leq W$):

\begin{definition}
For any D-string $\hat{S}$, its $i^{th}$ width (for $1\leq i \leq W$) $\hat{S}[i]$ is the letter, or the set of alternative letters, that can appear in $T[i]$ for string $T\in \hat{T}$. 
For any D-string $\hat{S}$, we denote by $\hat{S}[i_1,i_2]$ the \emph{D-substring} of $\hat{S}$ that starts at width $i_1$ and ends at width $i_2$ with $1\leq i_1\leq i_2\leq W$.
\end{definition}

We will also use the notation $s_i$ for both $|\hat{S}\{i\}|$ and $|\hat{S}[i]|$.


\begin{example}
The D-string $\hat{T}=\texttt{GCA[AT/CG]C[G/T]GG[TA/AA/AT]TT}$
 has length $n\!=\!11$, size $N\!=\!20$, and width $W\!=\!13$. 
 We have that at width, say, $3$ there is a solid position $\texttt{A}=\hat{S}[3]$, while at width $4$ there is $\texttt{[A/C]}=\hat{S}[4]$. Furthermore, the D-substring $\hat{S}[3,7]$ is $\texttt{A[AT/CG]C[G/T]}$. Here below we show positions and widths for the D-string $\texttt{CA[AT/CG]C[G/T]GG[TA/AA]T}$ having length 9, width 11, and size 16.
\end{example}

\noindent
\begin{tabular}[width=1.0\textwidth]{rccccccccccccccccccccccccc}
D-string &$\tt{C}$&$\tt{A}$&$\texttt{[A}$&$\texttt{T}$&/&$\texttt{C}$&$\texttt{G]}$&$\texttt{C}$&$\texttt{[G}$&/&$\texttt{T]}$&$\texttt{G}$&$\texttt{G}$&$\texttt{[T}$&$\texttt{A}$&/&$\texttt{A}$&$\texttt{A]}$&\texttt{T}\\
width &1&2&3&4&&3&4&5&6&&6&7&8&9&10&&9&10&11\\
position&1&2&&&3&&&4&&5&&6&7&&&8&&&9
\end{tabular}

\subsection{Partial Order Alignments}
\label{ssec:poa}

The use of Partial Order (graphs) for sequence Alignments (POA) was introduced in \cite{lee_multiple_2002} with the purpose of improving the iterative step of progressive (\cite{pMSA}) multiple sequence alignments (MSA), that is the step where a new sequence is added to an MSA (\emph{starting MSA}). Traditionally, this was performed by first reducing the starting MSA to a linear profile, and then aligning it to the new sequence in order to obtain a new MSA (\emph{resulting MSA}), with the downside that the reduction of the starting MSA to a linear profile might carry and propagate alignment errors. The idea behind POA is to perform a direct pairwise dynamic programming alignment between the new sequence and a suitable graph representation of the starting MSA (thus replacing the somewhat lossy linear profile representation), with the outcome of a guarantee that, in the resulting MSA, the optimal alignment of the new sequence with respect to each one of the sequences contributing to the starting MSA is maintained. Such graph representation, the \emph{partial order graph}, basically replaces the linear MSA profile with a DAG whose edges represent a partial order between the letters of the MSA to be used as an alternative to the traditional total order of the rows of an alignment.

\subsection{Wave Front Alignments with affine gap penalty}
\label{ssec:wfa}

In \cite{Myers86}, Gene Myers introduced the linear time and space dynamic programming (DP) alignment for similar strings: when aligning two strings of size $N$ whose distance is known to be upper-bounded by $D$, rather than computing the whole DP table of size $N^2$, only a stripe of $\cal{O}{D}$ diagonals are computed, as their similarity guarantees their optimal alignment to lay therein. The result is a $\cal{O}(DN)$ algorithm replacing the quadratic one. \\
The \emph{Affine Gap Penalty} function to score the cost of gaps in alignments evaluates a series of $k$ consecutive gaps as $w(k)=o+k\cdot e$ (where $o$ is the cost of opening the gap, and $e$ that of extending it), rather than the sum of $k$ single gap's costs. Using such gap penalty score is almost mandatory in genomic sequences analysis as this forces the optimal alignment to detect and highlight INDELs variations that typically involve several consecutive nucleotides. In order to restore the optimal substructure of the alignment problem when using affine gap penalty score funcion, the DP algorithm requires three alignment tables instead of one: matrices $I,D,M$ that store the score of the best alignments ending - respectively - with an Insertions, a Deletions, or a (Mis)Match.\\
In \cite{marco-sola_fast_2020}, the linear time and space optimization was extended to the exact computation of an optimal pairwise alignment of strings using the affine gap penalty score function. Roughly, for the alignment of two strings of size $N$ and $M$, the three matrices $I,D,M$ of size $N\times M$ are replaced by \emph{wave front} records $I_{d,k},D_{d,k},M_{d,k}$ that store, per each score/distance $d$ and diagonal $k$, the furthest-reaching offset in $k$ that scores $d$. The dynamic programming recurrence is then operated on increasing values of $d$ and adjacent diagonals $k$, up to the final distance.

\section{The DSA Algorithm}
\label{sec:algo}

Let us first formally state the problem we solve:

\defproblem{String Alignment to \ds (StoDS)}{A D-string $\hat{T}$ of width $W$ and size $N$, a pattern $P$ of length $m$, and penalty scores $a,x,o,e$.}{An optimal alignment between $P$ and $\hat{T}$ using scores: $a$ for match, $x$ for mismatch, $o$ for gap opening, and $e$ for gap extension.}

In the problem statement above, by optimal alignment we mean the alignment between $P$ and a string $T\in \hat{T}$ (see Section~\ref{ssec:ds}) that minimizes the distance computed by scoring $a$ for a match, $x$ for a mismatch, and gap affine penalty function with $o$ for gap opening and $e$ for gap extension.

In this section we describe our algorithm DSA that optimally solves \textsc{StoDS}. 
We start with Section~\ref{ssec:poa4ds} where we show how we adapted the \emph{Partial Order Alignment} (POA) framework to work with alignment to D-strings. Then, in Section~\ref{ssec:wfa4ds} we describe our extension of \emph{Wave Front Alignment} (WFA) to D-strings with and show how we merge the two techniques into our algorithm DSA and analyse its complexity.

\subsection{Partial Order Alignment with \ds}
\label{ssec:poa4ds}

\begin{figure}
\centering
\includegraphics[width=0.5\textwidth]{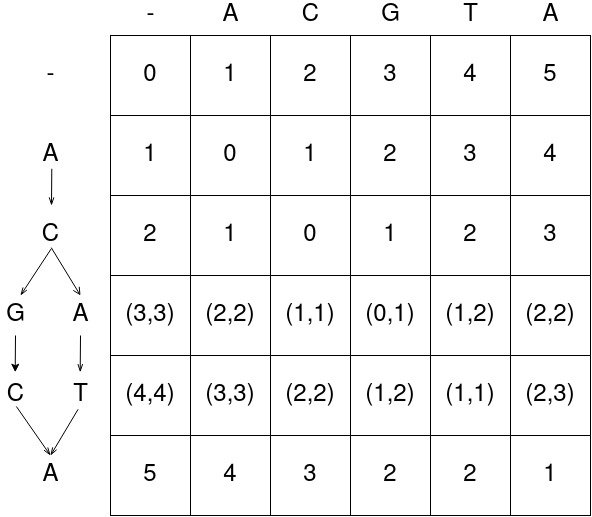}
\caption{Dynamic Programming table for the Alignment of the string $\texttt{ACGTA}$ against the D-string $\texttt{AC[GC/AT]A}$ partially ordered as shown by oriented edges, and using scores: 0 for match, 1 for mismatch, and 2 for gap.}
\label{fig:DPtable}
\end{figure}

In this section, we show how to use partial order alignment to be able to perform base level alignments with D-strings.
Without loss of generality, for the sake of simplifying the notation, within this section we will not use the gap affine penalty score (whose implementation with \ds will be described in the next section), but rather account a cost $g$ to any gap.
Given a D-string $\hat{T}$ of width $W$ and pattern $P$ of length $m$, we build a $(W+1)\times (m+1)$ DP table $M$ having a row $i$ per each $i^{th}$ width of $\hat{T}$, and a column per each letter of $P$ (plus the usual first row and first column for their empty prefixes), and that will store in $M[i,j]$ the best score of aligning $P[1,j]$ to $\hat{T}[1,i]$ (an example is shown in Figure~\ref{fig:DPtable}).  When a row falls within a non-solid position having $s$ strings of length $\ell$, then the row is not associated with a simple letter, but rather with $s$ of them, and this will be the case for $\ell$ consecutive rows.
The entries of these rows will not contain a single score, but rather a tuple of size $s$: if row $i$ has letters $a_{i,1}, \dots, a_{i,h}, \dots , a_{i,s}$, then $M[i,j]$ is a tuple of $s$ values $<M[i,j]^1, \dots ,M[i,j]^h,\dots ,M[i,j]^s>$ where, for each $1\leq h\leq s$, a match is accounted if and only if $a_{i,h}=P[j]$.\\
On top of this tuple representation, a partial order is assumed for the $N$ letters of the D-string: within solid positions, the usual order of sequences applies; in a degenerate position, instead, distinct letters in the same tuple are not comparable, and nor is letter $a_{i,h}$ of tuple $a_{i,1}, \dots, a_{i,h}, \dots , a_{i,s_i}$ at row $i$ comparable with the other letters of all the $\ell$ adjacent rows composing the same degenerate position, except for the $\ell$ other letters $a_{i',h}$ that are also the $h^{th}$ letter in their tuples. Within all comparable letters, the order corresponds to that of the rows of $M$.

We use this partial order to drive the alignment along the D-string: wherever the traditional dynamic programming alignment algorithm refers to the previous row and/or previous column, here we refer only to entries that - instead - are preceding according to the partial order defined above.

\begin{example}
For the D-string $\texttt{AC[GC/AT]A}$ the partial order is defined by the graph shown on the letters at the rows of the DP table of Fogure~\ref{fig:DPtable}.
\end{example}

We now formalize with dynamic programming recurrence relations the way we apply the ideas sketched above. For any two letters $x,y\in \Sigma$, we  define a function $m(x,y)$ as equal to $a$ if $x=y$, and equal to $x$ otherwise. Denoting with $s_i$ the size of the tuple at row $i$, we compute $M[i,j]$ for all $1\leq j \leq m$ distinguishing the following cases:

\begin{description}
\item[$s_{i-1}=s_i=1$.] In this case no degeneracy is encountered and the traditional dynamic programming alignment framework applies.
\item[$s_{i-1}=1$ and $s_i=s>1$.] In this case, row $i$ corresponds to the opening of a degenerate position of $s_i$ strings of length $\ell$, this row (like the next $\ell-1$ rows) represents $s_i$ alternative letters $a_{i,1}, \dots, a_{i,s_i}$ and each entry $M[i,j]$ will contain tuples of size $s_i$ computed as follows for $1\leq h \leq s_i$:
\begin{equation}
M[i,j]^h=min
 \begin{cases}
        M[i-1,j-1]+m(P[j],a_{i,h}) \\
        M[i-1,j]+ g\\
        M[i,j-1]^h+ g\\
      \end{cases}
\end{equation}

\item[$s_{i-1}=s_i=s>1$.] In this case, row $i$ is inside the same degenerate position of the previous row\footnote{Should rows $i-1$ and $i$ fall into distinct degenerate positions that have the same size only by chance, this case would be managed as in the last case below.}, this row still represents $s$ alternative letters $a_{i,1}, \dots, a_{i,s_i}$ and each entry $M[i,j]$ will contain tuples of size $s_i$ computed as follows for $1\leq h \leq s_i$:
\begin{equation}
M[i,j]^h=min
 \begin{cases}
        M[i-1,j-1]^h+m(P[j],a_{i,h}) \\
        M[i-1,j]^h+ g\\
        M[i,j-1]^h+ g\\
      \end{cases}
\end{equation}

\item[$s_{i-1}=s>1$ and $s_i=1$.] In this case, row $i-1$ was the last letter of a degenerate position, and row $i$ is a solid position representing a single letter $\hat{T}[i]$. The entry $M[i,j]$ will contain a value computed as follows:
\begin{equation}
M[i,j]=min
 \begin{cases}
        M[i-1,j-1]^1+m(P[j],\hat{T}[i]) \\
        \dots \hspace{1cm} \dots \hspace{1cm} \dots  \\
        M[i-1,j-1]^{s_{i-1}}+m(P[j],\hat{T}[i]) \\
        M[i-1,j]^1+ g\\
        \dots \hspace{1cm} \dots \\
         M[i-1,j]^{s_{i-1}}+ g\\
        M[i,j-1]+ g\\
      \end{cases}
\end{equation}

\item[$s_{i-1}>1$ and $s_i=s>1$  but not necessarily $s_{i-1}=s_i$.] In this case, row $i-1$ was the last letter of a degenerate position and row $i$ is the first of a new different degenerate position. This row represents $s$ alternative letters $a_{i,1}, \dots, a_{i,s_i}$ and each entry $M[i,j]$ will contain tuples of size $s_i$ computed as follows for $1\leq h \leq s_i$:
\begin{equation}
M[i,j]^h=min
 \begin{cases}
        M[i-1,j-1]^1+m(P[j],a_{i,h}) \\
        \dots \hspace{1cm} \dots \hspace{1cm} \dots  \\
        M[i-1,j-1]^{s_{i-1}}+m(P[j],a_{i,h}) \\
        M[i-1,j]^1+ g\\
        \dots \hspace{1cm} \dots \\
         M[i-1,j]^{s_{i-1}}+ g\\
        M[i,j-1]^h+ g\\
      \end{cases}
\end{equation}
\end{description}

For example, Figure~\ref{fig:DPtable} shows the DP table for the alignment of the D-string $\texttt{AC[GC/AT]A}$ against $\texttt{ACGTA}$.

\subsection{Wave Front Alignment with \ds}
\label{ssec:wfa4ds}

In this section we show how to perform Wave Front Alignment of Partially Ordered D-Strings with affine gap penalty score in almost linear time using a \ds customization of Wave Front Alignment.

\begin{table}[htbp]
\caption[Notation]{\label{tbl:notation}Notation}
\centering
\begin{tabular}{ll}
Symbol & Description\\
\hline
a & score of match\\
x & score of mismatch\\
o & score of gap opening\\
e & score of gap extending\\
d & current distance (it will start from $0$ and reach the solution of the \textsc{StoDS} problem)\\
P & query pattern: \(P = P_0P_1...P_m\) ($m=\Theta(W)$)\\
$\hat{T}$ & D-string \(\hat{T}=\hat{T}[0]\hat{T}[1]...\hat{T}[W]\)\\
$WF_d$ & the wavefront $WF_d$ for $d$\\
\(\widetilde{M}_d, \widetilde{I}_d, \widetilde{D}_d\) & components of $WF_d$ that will be defined for each diagonal $k$ in $WF_d$. \\
\(\widetilde{M}_{d,k}, \widetilde{I}_{d,k}, \widetilde{D}_{d,k}\) & contain the values offset$_{d,k}$ for each diagonal $k$ in $WF_d$. \\
$lo_d, hi_d$ & define the width of $WF_d$\\
offset$_{d,k}$ & defines for each diagonal $k$ in $WF_d$ the furthest reaching points with distance $d$.\\
abandoned$_{d,k}$ & boolean telling whether furthest point has been reached in diagonal $k$ for distance $d$.  \\ 
\end{tabular}
\end{table}

Similarly to \cite{marco-sola_fast_2020}, we replace the three affine gap penalty dynamic programming matrices $I,D,M$ of size $W\times m$ ($W$ being the width of $\hat{T}$ and $m$ the length  of $P$) with \emph{wave front} records $\widetilde{I}_d$, $\widetilde{D}_d$, $\widetilde{M}_d$ starting with an initial distance $d=0$. The value of $d$ can thus only increase along the computation, and for each partial score $d$, the records $\widetilde{I}_d$ (resp. $\widetilde{D}_d$, $\widetilde{M}_d$) store the following values: $lo_{d}, hi_{d}$ and a tuple offset$_{d,k}$ per each diagonal $k$ in the range defined by $lo_{d}$ and $hi_{d}$.
Indeed, the values $lo_{d}, hi_{d}$ define $WF_d$: the set of diagonals that allow to reach alignment score $d$. 
Per each partial distance $d$ and diagonal $k$, offset$_{d,k}$ stores the furthest-reaching offset in diagonal $k$ that scores $d$. Initially, $lo_{d}= hi_{d}=0$ and the only single starting diagonal is the main diagonal $k=W-m$.

Table~\ref{tbl:notation} summarizes the symbols and data (structure) we are using.

When using $WF_d$ with D-texts, $\widetilde{I}_{d, k},\widetilde{D}_{d, k},\widetilde{M}_{d, k}$ are in general tuples of size $s$ (the number of variants).
Observe that ($d$, $k$, and) offset$_{d,k}$, identify a specific diagonal and how far you can go down along it keeping score $d$, and therefore $k$ and offset$_{d,k}$ actually precisely define an entry in what would have been the matrices $D,I$ and $M$. Let us name $u$ the row and $v$ the column of the entry defined by offset$_{d,k}$.
Now, if $\hat{T}[u]$ is a single/solid letter, then the offset of $\widetilde{I}_{d},\widetilde{D}_{d},\widetilde{M}_{d}$ is a single value, and else it is a tuple of size $s_u$: a wave front per each letter $\hat{T}[u]$.\\
Starting with initial distance $d=0$ and the sole central diagonal, we compute the wavefront for growing values of $d$ up the final result. The value of  $d$ is increased, and new offsets are computed, when the borders of the wave front have been reached: no further match is possible, and either a mismatch score $x$ must be accounted (and new offset computed along the same diagonal $k$ for distance $d+x$), or a gap must be accounted and therefore new diagonals must be explored (and new offset computed for distance that has increased according to the gap penalty function).\\
Let us now formalize the Wave Front Alignment for \ds with the recurrence relations that show how to update, for each diagonal $k$, the offsets information, 
We will use the generic notation offset$_{d,k}$ as well as the more specific notation offset$^{D}_{d,k}$ (resp. offset$^{I}_{d,k}$, offset$^{M}_{d,k}$) to  mean the offsets in $\widetilde{D}_d$ (resp. $\widetilde{I}_d$, $\widetilde{M}_d$).\\
The formulae below are basically those of \cite{marco-sola_fast_2020} with the tuple information added.
When computing $\widetilde{I}_{d,k}$ we are assuming that the last event has been an insertion, in which case the previous event could either have been (i) a (mis)match, in which case a new gap is being opened, or (ii) another insertion, in which case an existing gap is being extended. In case (i) the new offset is that of $\widetilde{M}_{d-o-e}$ because the distance is increased by $o+e$, while in case (ii) the new offset is that of $\widetilde{I}_{d-e}$ because the distance is only increased by the gap extension $e$. In both cases, the new offset has to be picked from the preceding diagonal $k-1$. The following formula formalizes this, describing how to compute the new offsets for $\widetilde{I}_{d,k}$, assuming the general case in which the entry on which we do the recurrence contains a tuple of size $s$:

\begin{equation}
\label{eq:I-d-k}
\begin{array}{ll}
      offset^I_{d,k} = 1+ max
      \begin{cases}
        \underset{\texttt{in tuple of size s}}{\max}(offset^M_{d-o-e,k-1}) & \text{(Open insertion)} \\
        \underset{\texttt{in tuple of size s}}{\max}(offset^I_{d-e,k-1})  & \text{(Extend insertion)}
      \end{cases}
 \end{array}
\end{equation}

Dually, $\widetilde{D}_{d,k}$ is computed assuming that the last event has been a deletion, in which case the previous event could either have been a (mis)match, in which case a new gap is being opened, or another deletion, in which case an existing gap is being extended. The new offset is then either in $\widetilde{M}_{d-o-e}$ or in  $\widetilde{D}_{d-e}$, and in both cases the new offset has to be picked from the next diagonal $k-1$. The following formula formalizes this, describing how to compute the new offsets for $\widetilde{D}_{d,k}$, assuming the general case in which the entry on which we do the recurrence contains a tuple of size $s$:

\begin{equation}
\label{eq:D-d-k}
\begin{array}{ll}
      offset^D_{d,k} = max
      \begin{cases}
        \underset{\texttt{in tuple of size s}}{\max}(offset^M_{d-o-e,k+1}) & \text{(Open deletion)} \\
        \underset{\texttt{in tuple of size s}}{\max}(offset^D_{d-e,k+1}) & \text{(Extend deletion)}
      \end{cases}
 \end{array}
\end{equation}

Dually, $\widetilde{M}_{d,k}$ is computed assuming that the last event has been a (mis)match, in which case we stay in the same diagonal $k$, and the previous event could either have been another (mis)match, or an insertion or deletion. The new offset is then either in $\widetilde{M}_{d-x}$ or in  $\widetilde{D}_{d}$ or $\widetilde{I}_{d}$. The following formula formalizes this, describing how to compute the new offsets for $\widetilde{M}_{d,k}$, assuming the general case in which the entry on which we do the recurrence contains a tuple of size $s$:

\begin{equation}
\label{eq:M-d-k}
\begin{array}{ll}
      offset^M_{d,k} = max
      \begin{cases}
        offset^I_{d, k} & \text{(Insertion)} \\
        offset^D_{d, k} & \text{(Deletion)} \\
        \underset{\texttt{in tuple of size s}}{\max}(offset^M_{d-x, k}) + 1 & \text{(Mismatch)} \\
      \end{cases}
\end{array}
\end{equation}

In the formula for $\widetilde{M}_{d,k}$, the case of a match is not taken into account because offsets are recomputed only when a match is not available.

Notice that, among possible values, the maximum is always sought here because the distance is a fixed parameter $d$ in $\widetilde{M}_{d,k}, \widetilde{I}_{d,k}, \widetilde{D}_{d,k}$, and with that one wants to go as far as possible along $k$ maintaining that distance (that can only increase otherwise), therefore eventually minimizing the final $d$ whose $WF_d$ will reach the bottom right final entry of the matrix.

In all formulae above, whenever incurring in a gap and increasing or decreasing the diagonal number, then this has to be intended as updating $lo_d$ or $hi_d$.

The pseudocode of DWF\_ALIGN is shown in Algorithm~\ref{dwf_align} (structured as in \cite{marco-sola_fast_2020} but modified to be adapted to \ds and their tuples based partial order alignment realization). The function DWF\_EXTEND (Algorithm~\ref{dwf_extend}) is called to extend the wavefront extends $\widetilde{M}_{d,k}$ for all diagonals $k$ between $lo_d$ and $hi_d$, and it calls the function $\lambda$ (Algorithm~\ref{match_lambda}), which is the core of \tool. 
This function is where the offsets are actually increased. When this happens, then $\lambda$:\\
(i) Updates the width of the offset tuple with that of the current row to account for the case in which we extend a wavefront from an index $i$ by 1 into an index $i+1$ such that $|\widehat{T}[i]| \neq |\widehat{T}[i+1]|$ (the size of the tuple changes), that happens when the wavefront has been extended into a degenerate letter or out of a degenerate letter. In this case, we need to increase or reduce the length of offsets to correspond to the number of variants at $\widehat{T}[i+1]$. We do this by selecting $max(\widehat{T}[i])$ and setting it as the offset for each variant at $\widehat{T}[i+1]$. \\
(ii) Picking the maximum from previous row does not incur into an artefact match\footnote{With artefact match we mean to, say, align exactly pattern $P=ACGT$ to $\widehat{T}=A\{TC,GA\}T$ matching the two letters $CG$ that belong to distinct variants.} thanks to the boolean vector $abandoned$ (if that part of a variant cannot be extended any further, then we mark as abandoned the extension of the wavefront for the given score) that will allow to know along which variant we can extend the wavefront, and will also be used by the traceback function to compute the correct alignment

\begin{algorithm}
  \caption{}\label{dwf_align}
  \begin{algorithmic}
    \Function{DWF\_ALIGN}{$\widehat{T}, P, \lambda$}

       \State $W\gets |w(\widehat{T})|$\Comment{width of the D-Text}
       \State $A_k\gets |\widehat{T}| - |P|$\Comment{central diagonal}
      \State $end \gets max\{W,|P|\}$ \Comment{end of alignment index}
       \State $d \gets 0$; $s_0 \gets |\widehat{T}[0]|$ \Comment{initial edit distance and width}
       
       \State offset$^{M}_{0,A_k}\gets <0, ...,0>$ \Comment{tuple of length $s_0$ initialized to 0s }
       \State abandoned$^{M}_{0,A_k}\gets <false, ...,false>$ \Comment{tuple of length $s_0$ initialized to false}
       \While{true}
         \LComment{extend the M-wavefront}
         \State \Call{DWF\_EXTEND}{$\widetilde{M}, \lambda$, d}

        \State $f \gets \texttt{max in tuple }offsets^M_{d,A_k}$ 
         \If{$f \geq  end$} 
               break;
         \EndIf  
        \Comment{ reached end of alignment}
         \State $d \gets d+1$
         \Comment{increase current distance}
         \State \Call{DWF\_NEXT}{$\widetilde{M}, \widetilde{I}, \widetilde{D}$, d} 
         \Comment{compute the bounds of the next wavefront}

       \EndWhile

       \State  $A \gets \Call{DWF\_TRACEBACK}{\widetilde{M}, \widetilde{I}, \widetilde{D}, d}$

       \LComment{return the edit distance and the alignment}
       \State \textbf{return} $(d, A)$

    \EndFunction
\end{algorithmic}
\end{algorithm}

\begin{algorithm}
  \caption{}\label{dwf_extend}
  \begin{algorithmic}
    \Function{DWF\_EXTEND}{$\widetilde{M}, \lambda$, d}

       \For{$k$ {\bf from} $\widetilde{M}_{d}^{lo}$\ \textbf{to}\ $\widetilde{M}_{d}^{hi}$}
         \State $ab \gets abandoned^{M}_{d,k}$
         \Comment{tuple of abandoned variants}
         \State $of \gets offsets^{M}_{d,k}$
         \Comment{tuple of offsets}
         \State $f \gets \texttt{max in tuple} of$
         \Comment{index in $\widehat{T}$}
         \State $u \gets f$
         \Comment{index in $P$}
         \State $v \gets f-k$
         \Comment{extend $WF_d$}
         \While{\Call{$\lambda$}{$v, u, of, ab$}} 
         \EndWhile
       \EndFor

    \EndFunction
\end{algorithmic}
\end{algorithm}

\begin{algorithm}
  \caption{}\label{match_lambda}
  \begin{algorithmic}
    \Function{$\lambda$}{v, u, offsets, abandoned}

       \If{$u<0$\ \textbf{or}\ $v < 0$\ \textbf{or}\ $v \geq |P|$\ \textbf{or}\ $u  \geq W$}
        \textbf{return} $false$\
       \EndIf

       \State $s_{prev} \gets length(offsets)$
        \Comment{size of previous tuple}
       \State $s_{curr} \gets |\widehat{T}[u]|$
        \Comment{size of current tuple}

       \LComment{we are stepping into or out of a degenerate letter}
       \If{$s_{curr} \neq s_{prev}$}
          \LComment{furthest offset}
          \State $f \gets \texttt{max in tuple } offsets$

          \LComment{create tuple of size $s_{curr}$ with furthest offset}
          \State $offsets \gets <f, ... ,f>$

          \LComment{create tuple of $s_{curr}$ size with false}
          \State $abandoned \gets <false, ... ,false>$
       \EndIf
       \State $extend \gets false$
       \State $previously\_extended \gets false$
       \LComment{attempt to extend each variant by 1}
       \For{$s'$ {\bf from} $0$\ \textbf{to}\ $s_{curr}$}
          \If{$\widehat{T}[h][s'] =  P[v] \textbf{and} abandoned[s'] = false$}
            \State $offsets[s'] \gets offsets[s'] + 1$
            \State $extend \gets true$

            \If{$previously\_extended = false$}
              \State $v \gets v+1$
              \State $u \gets h+1$
              \State $previously\_extended \gets true$
            \EndIf
          \EndIf
          \If{$\widehat{T}[u][s'] \neq  P[v]$}
            \State $abandoned[s'] \gets true$
          \EndIf
       \EndFor
       \State \textbf{return} $extend$\
    \EndFunction
\end{algorithmic}
\end{algorithm}

The final result we have obtained is an exact algorithm that computes the edit distance (or any distance accounting for insertions, deletions, and substitutions with any desired score, including affine gap penalty score), between a string and a D-string in time and space proportional to $D\cdot N$, where $N$ id the size of the D-string, and $D$ is the computed distance.

We remark that both the optimization problem we addressed and the dynamic programming solution we suggested can naturally make use of weights to be added to input data to be included in the distance function to be optimised. These could be confidence value for bases as well as multiplicity values in the D-string (accounting for variants that are more frequent than others). 

We implemented our algorithm in a prototype tool dsa that computes the edit distance and an optimal alignment (thus keeping all the information needed for the tracing back the optimal path). In the next section we show some first comparative experimental validation of the performances of \tool.

\section{Experimental validation}
\label{sec:exp}

\subsection{State of the art}

To the best of our knowledge, there is no published software tool implementing an algorithm specifically designed for a global alignment of a linear string and a D-text.
Some algorithm are designed for \emph{semi}global alignment, that is with the pattern being (substantially) shorter than the D-string (e.g. \cite{garrison2018,vg2021,cislak_sopang_2018,cislak_sopang_2020,vargas}); some of them and others are designed for more general purposes, such as aligning a linear string to a graph structure that generalizes \ds (e.g.\cite{garrison2018,vg2021,vargas,DBLP:journals/bioinformatics/RautiainenMM19,graphaligner,minigraph}). Some of these and other tools are not exact and hence do not guarantee to find the optimal alignment or solution (e.g \cite{graphaligner,DBLP:journals/bioinformatics/RautiainenMM19,minigraph,minimap,minimap2,minimap3}). Finally, some algorithms that basically solve a very similar problem, actually do not compute any distance or alignment, but rather output a consensus string that suitably merges the input strings (e.g. \cite{abpoa,astarixrecomb20,astarixbiarxiv22,astarixrecomb22}).\\

The Variation Graph Toolkit \textsc{vg} suite \cite{garrison2018,vg2021} contains a tool that can solve \textsc{StoDS} aligning a string to a more general data structure than \ds (Variation Graphs indeed), but this is specifically designed to map reads that are (much) shorter that the graph/text, rather than making a global alignment, and therefore a comparison would not be fair. The same holds for \textsc{sopang} and its extension \textsc{sopang2} (\cite{cislak_sopang_2018,cislak_sopang_2020}): they are designed for \emph{short} reads mapping on ED-strings, and moreover they only detect exact matches without allowing gaps nor mismatches. Also the tool \textsc{vargas} (\cite{vargas}) works with short reads.\\
We thus considered the tool abPOA (\cite{abpoa}), a C library tool to align a sequence to a directed acyclic graph that also uses partial order alignment and supports global alignment, but we could not run in on data of length $100,000$b (accuracy comparison would not have been possible anyhow, as abPOA does not output an alignment nor a distance, but rather a consensus string. The same (could not run on length $100,000$b strings and output is a consensus string) holds for the tool Astarix (\cite{astarixrecomb20,astarixbiarxiv22,astarixrecomb22}), which addresses the string to graph alignment by solving a shortest path problem on a suitably designed graph representation of input data.\\

We therefore compared the performances of \textsc{DSA} with those of  \textsc{GraphAligner} (\cite{graphaligner,DBLP:journals/bioinformatics/RautiainenMM19}) and \textsc{Minigraph} (\cite{minigraph,minimap,minimap2,minimap3}) for solving \textsc{StoDS}. Both of them are designed to align strings on a more general graph structure that \ds and therefore the comparison we show below should be intended as a validation of the performances of \tool in solving \textsc{StoDS}, and not as \tool being in general a better tool than any of the other two.\\
\textsc{GraphAligner} is designed to align reads to genome graphs as a key to many applications, including error correction, genome assembly, and genotyping of variants in a pangenome graph. In aligning long reads to genome graphs, \textsc{GraphAligner} was 13x faster and used 3x less memory when compared in \cite{graphaligner} to the state-of-the-art tools. As for accuracy, when employing \textsc{GraphAligner} for error correction, in \cite{graphaligner} it turned out more than twice as accurate and over 12x faster than extant tools. \textsc{GraphAligner} uses a seed and extend method and the bitvector alignment extension algorithm of \cite{DBLP:journals/bioinformatics/RautiainenMM19}.\\
\textsc{Minigraph} is a sequence-to-graph mapper and graph constructor. It aligns a query sequence against a sequence graph using a gap affine penalty score. \textsc{Minigraph} is a well maintained and highly optimized software tool that uses minimizers to find strong colinear chains as starting point to build the alignment (\cite{minigraph,minimap,minimap2,minimap3}). \\


\subsection{Experiments setup}

We randomly generated a D-string of width $W=100,000$b by first generating a random string of length $W$ on alphabet $\{A,C,G,T\}$, and then inserting\footnote{The insertion was done forcing the width to remain $W$} therein $g$ (input parameter to our random generator of \ds given as a percentage of $W$) degenerate non-solid positions as follows: for each such position $p$ we pick at random a value for its size $s_p$ between $1$ and $S$ (another input parameter), and a randomly chosen length $\ell_p$ between $1$ and $L$ (input parameter again).\\
We generated \ds using width $W=100,000$b in all tests, degeneracy frequencies values $g=1\%, 10\%, 20\%$, maximum variance values $S=2,5$, and maximum variant lengths $L=1,4$. As a consequence the width of tested \ds will always be $W=100,000$b, while its total size $N$ will be greater than $W$ and depending from input parameters $g,S,L$.\\
To this purpose, we implemented the random \ds generator {\sc simed}, which is publicly available at \url{https://github.com/urbanslug/simed/}.\\
From the obtained synthetic D-string $\hat{T}$, we extracted a ground truth exact pattern $P_0$ of size $W$ (that is, a string $P\in \hat{T}$ that thus matches $\hat{T}$ with distance $0$), and  we (possibly) modified $P_0$ into the actual input query $P$ with different possible divergences using real {\rm .vcf} files.\\ 
The divergences we tested were to insert no divergence at all, or $0,1\%$ or $1\%$ SNPs, or $0,1\%$ INDELs (percentages are on $W$).
Hence, the size $m=|P|$ of the query string will be in $\Theta(W)$ and so will the distance $d$ between $P$ and $\hat{T}$.

\subsection{Preliminary Results}

All tests were ran on a laptop (single threaded) Intel® Core™ i7-11800H × 16 with 16.0 GiB RAM. Space and time was measured using \texttt{/usr/bin/time -f"\%S\textbackslash{}t\%M"} to extract system time (seconds) and maximum resident set size (kbytes). In all tests below we used alignment scores $a=0, x=1, o=2, e=1$.\\

In all experiments, for all tools time was below $0.1$ seconds and therefore negligible and not reported case per case. We do report, instead the memory peak (weakness of the current prototype of \tool: as we mention in further work session, we plan to improve that), the number of mistakes in the alignment found (only for the cases with no INDELs because there the alignment is affected by differences in penalty scores), and finally the last column shows the number of detected events on optimal alignment: number M of matches, X for mismatches, I for insertions, and D for deletions. When the pattern divergence is only SNPs, then for a correct alignment it must be $I=D=0$, $X$ approximately equal to the number of SNPs (except where by chance the divergence will not change the DNA base), and $M=W-X$. For the experiments involving INDELs divergence in the pattern, we also report the number $G$ of gaps that are opened: with $W=100,000$b and $0.1 \% INDELs$, in a correct alignment it must be $G=100$.

Table~\ref{res:1-2-1} below shows results with a D-string of size $N=101,000$ generated with lowly frequent low degeneracy (only $g=1\%$ of positions are degenerate, with at most $S=2$ variants of length $L=1$), and with four different pattern divergence: no divergence, $0.1\%$ of SNPs, $1\%$ of SNPs, and $0.1\%$ of INDELs.\\

\begin{table}[h]
\caption{\label{res:1-2-1}Performances with D-string of size $N=101,000$ with degeneracy $g=1\%$, $S=2$, $L=1$.}
\begin{tabular}{lllcl}
Tool & Pattern & Peak memory & Total Error & Alignment\\
  & divergence  & (kbytes) & (absolute) & events\\
\hline
\tool & none & 3162 & 0 & 100,000M 0X 0I 0D\\
\textsc{GraphAligner} & "  & 6244 & 0 & 100,000M 0X 0I 0D\\
\textsc{Minigraph} & "  & 5607 & 9 & 99,991M 9X 0I 0D\\
\hline
\tool & 0.1 \% SNPs  & 42209 & 0 & 99902M 98X 0I 0D\\
\textsc{GraphAligner} & "  & 23972 & 1 & 99903M 97X 0I 0D\\
\textsc{Minigraph} & "  & 8058 & 2 & 99891M 100X 0I 0D\\
\hline
\tool & 1 \% SNPs & 474963 & 0 & 99003M 997X 0I 0D\\
\textsc{GraphAligner} & "  & 23960 & 10 & 99001M 995X 4I 4D\\
\textsc{Minigraph} & "  & 7928 & 3 & 98991M 1000X 0I 0D\\
\hline
\tool & 0.1 \% INDELs  & 99322   & 0 & 99865M 0X 106I 133D 100G\\
\textsc{GraphAligner} & "  & 23973 & - & 99757M 111X 105I 132D 140G\\ 
\textsc{Minigraph} & "  & 8162 & - & 99865M 0X 106I 133D 100G\\
\end{tabular}
\end{table}

Table~\ref{res:1-5-4} below shows results with a D-string of size $N=106,147$ generated with lowly frequent high degeneracy (only $g=1\%$ of positions are degenerate, with up to $S=5$ variants of length up to $L=4$), and with four different pattern divergence: no divergence, $0.1\%$ of SNPs, $1\%$ of SNPs, and $0.1\%$ of INDELs. \\

\begin{table}[h]
\caption{\label{res:1-5-4}Performances with D-string of size $N=106,147$ with degeneracy $g=1\% , S=5, L=4$.}
\begin{tabular}{lllcl}
Tool & Pattern  & Peak memory & Total Error & Alignment\\
 & divergence  & (kbytes) & (absolute) & events\\
\hline
\tool & none  & 3188 & 0 & 100,000= 0X 0I 0D\\
\textsc{GraphAligner} & "  & 7418 & 0 & 100,000= 0X 0I 0D\\
\textsc{Minigraph}  & "  & 5585 & 11 & 99989= 0X 0I 0D\\
\hline
\tool & 0.1 \% SNPs  & 44096 & 0 & 99900= 100X 0I 0D\\
\textsc{GraphAligner} & "  & 26799 & 1 & 99901M 99X 0I 0D\\
\textsc{Minigraph}  & " & 8038 & 11 & 99889M 100X 0I 0D\\
\hline
\tool & 1 \% SNPs  & 475668 & 0 & 99007= 993X 0I 0D\\
\textsc{GraphAligner} & "  & 26682 & 6 & 99005M 993X 2I 2D\\
\textsc{Minigraph}  & "  & 7910 & 11 & 98989M 1000X 0I 0D\\
\hline
\tool & 0.1 \% INDELs  & 131311 & 0 & 99813M 0X 135I 174D 100G\\
\textsc{GraphAligner} & "  & 26816 & - & 99717M 109X 135I 174D 145G\\
\textsc{Minigraph}  & "  & 8213 & - & 99813M 1X 136I 175D 100G\\
\end{tabular}
\end{table}

Table~\ref{res:10-2-1} below shows results with a D-string of size $N=110,000$ generated with medium frequent low degeneracy ($g=10\%$ of positions are degenerate, with only up to $S=2$ variants of length $L=1$), and with four different pattern divergence: no divergence, $0.1\%$ of SNPs, $1\%$ of SNPs, and $0.1\%$ of INDELs. \\

\begin{table}[h]
\caption{\label{res:10-2-1}Performances with D-string of size $N=110,000$ with degeneracy $g=10\% , S=2, L=1$.}
\begin{tabular}{lllcl}
Tool & Pattern  & Peak memory & Total Error & Alignment\\
 & divergence  & (kbytes) & (absolute) & events\\
\hline
\tool & none & 3174 & 0 & 100,000M 0X 0I 0D\\
\textsc{GraphAligner} & " & 16405 & 0 & 100,000M 0X 0I 0D\\
\textsc{Minigraph} & "  & 5609 & 6 & 99994M 0X 0I 0D\\
\hline
\tool & 0.1 \% SNPs & 46942 & 0 & 99903M 97X 0I 0D\\
\textsc{GraphAligner} & "  & 50386 & 0 & 99903M 97X 0I 0D\\
\textsc{Minigraph} & "  & 8078 & 12 & 99894M 100X 0I 0D\\
\hline
\tool & 1 \% SNPs  & 456316 & 0 & 99022M 978X 0I 0D\\
\textsc{GraphAligner} & "  & 50250 & 12 & 99020M 976X 4I 4D\\
\textsc{Minigraph} & " & 7934 & 94 & 98947M 997X 0I 0D\\
\hline
\tool & 0.1 \% (INDELS) & 105904 & 0 & 99806M 0X 171I 183D 100G\\
\textsc{GraphAligner} & "  & 50223 & - & 99708M 109X 171I 183D 160G\\
\textsc{Minigraph} & "  & 8145 & - &  99806M 0X 171I 183D 100G\\
\end{tabular}
\end{table}

Table~\ref{res:10-5-4} below shows results with a D-string of size $N=160,327$ generated with medium frequent high degeneracy ($g=10\%$ of positions are degenerate, with up to $S=5$ variants of length up to $L=4$), and with four different pattern divergence: no divergence, $0.1\%$ of SNPs, $1\%$ of SNPs, and $0.1\%$ of INDELs. 

\begin{table}[h]
\caption{\label{res:10-5-4}Performances with D-string of size $N=160,327$ with degeneracy $g=10\% , S=5, L=4$.}
\begin{tabular}{lllcl}
Tool & Pattern  & Peak memory & Total Error & Alignment\\
 & divergence  & (kbytes) & (absolute) & events\\
\hline
\tool & none  & 3134 & 0 & 100,000= 0X 0I 0D\\
\textsc{GraphAligner} & "  & 27417 & 0 & 100,000= 0X 0I 0D\\
\textsc{Minigraph} & "  & 5536 & 4 & 99996= 0X 0I 0D\\
\hline
\tool & 0.1 \% SNPs  & 66656 & 0 & 99905= 95X 0I 0D\\
\textsc{GraphAligner} & "  & 76167 & 0 & 99905M 95X 0I 0D\\
\textsc{Minigraph} & "   & 8046 & 14 & 99896M 100X 0I 0D\\
\hline
\tool & 1 \% SNPs & 469600 & 0 & 99022= 978X 0I 0D\\
\textsc{GraphAligner} & "  & 76129 & 13 & 99024M 973X 3I 3D\\
\textsc{Minigraph} & "  & 7878 & 48 & 98804M 1000X 0I 0D\\
\hline
\tool & 0.1 \% (INDELS)  & 126051 & 0 & 99814M 0X 108I 186D 100G\\
\textsc{GraphAligner} & "  & 76626 & - & 99665M 147X 110I 188D 170G\\
\textsc{Minigraph} & "  & 8199 & - & 99804M 0X 108I 186D 100G\\
\end{tabular}
\end{table}

Table~\ref{res:20-2-1} below shows results with a D-string of size $N=120,000$ generated with frequent but low degeneracy ($g=20\%$ of positions are degenerate, with only up to $S=2$ variants of length $L=1$), and with four different pattern divergence: no divergence, $0.1\%$ of SNPs, $1\%$ of SNPs, and $0.1\%$ of INDELs. \\

\begin{table}
\caption{\label{res:20-2-1}Performances with D-string of size $N=120,000$ with degeneracy $g=20\% , S=2$ and $L=1$.}
\begin{tabular}{lllcl}
Tool & Pattern  & Peak memory & Total Error & Alignment\\
 & divergence  & (kbytes) & (absolute) & events\\
\hline
\tool & none &  3166 & 0 & 100000M 0X 0I 0D\\
\textsc{GraphAligner} & "  & 21224 & 0 & 100000M 0X 0I 0D\\
\textsc{Minigraph}  & "  & 5604 & 6 & 99994M 0X 0I 0D\\
\hline
\tool & 0.1 \% SNPs  & 53252 & 0 & 99901M 99X 0I 0D\\
\textsc{GraphAligner} & "  & 76291 & 2 & 99902M 98X 0I 0D\\
\textsc{Minigraph}  & "  & 8078 & 8 & 99894M 100X 0I 0D\\
\hline
\tool & 1 \% SNPs  & 436404 & 0 & 99048M 952X 0I 0D\\
\textsc{GraphAligner} & "  & 76002 & 6 & 99047M 951X 2I 2D\\
\textsc{Minigraph}  & "  & 7913 & 102 & 98994M 1000X 0I 0D\\
\hline
\tool & 0.1 \% (INDELS)  & 103666  & 0 & 99908M 0X 96I 88D 100G\\
\textsc{GraphAligner} & "  & 76305 & - & 99813M 91X 150I 96D 141G\\
\textsc{Minigraph}  & "  & 8132 & - & 99908M 0X 96I 88D 100G\\
\end{tabular}
\end{table}

In all tables, the first experiment is for an exact match: no divergence is introduced in the pattern, and an exact match is supposed to be found. As we can see above, \tool always does find the exact solution with 100000 matches, 0 mismatches, and 0 gaps (like \textsc{GraphAligner} does) taking less memory than \textsc{Minigraph} and much less than \textsc{GraphAligner}.\\
In the second and third experiments we introduced SNPs (in $0.1 \%$ and $1 \%$ of the positions, respectively): in all experiments, \tool is the only one which is exact (except in two cases where for $0.1 \%$ SNPs also \textsc{GraphAligner} is correct but with higher memory usage) at a cost of the highest memory consumption with $1 \%$ SNPs, and only higher than \textsc{Minigraph}'s for $0.1 \%$ SNPs (but therein \textsc{Minigraph} is always the least accurate).\\
Finally, with INDELs, we can see that the memory usage of \tool must be improved, but our prototype is always exact (and \textsc{Minigraph} almost is, as only in one case it detects a spurious mismatch). \textsc{GraphAligner} is almost always wrong which is expected given it is more general and does not aim to find an exact alignment.

\section*{Conclusions and further work}

As mentioned above, at the moment our implementation of \tool is just a promising prototype. We plan to improve its memory consumption using the ideas of \cite{EP22}. Also, we are working on improve its speed (as well as consequent memory use) by removing hopeless diagonals suffixes that are currently being kept in the range. \\
Also, we remark that with our dynamic programming method, it is very natural to add weights to letters and to use a sum of weights modification of the actual score as objective function, in order to account for possible useful metadata such as confidence level in the query string or D-strings bases, or abundance in the MSA represented by the D-string.

\bibliography{references}

\end{document}